\documentclass[sigconf, nonacm]{acmart}
\AtBeginDocument{%
  \providecommand\BibTeX{{%
    \normalfont B\kern-0.5em{\scshape i\kern-0.25em b}\kern-0.8em\TeX}}}

\setcopyright{acmcopyright}
\copyrightyear{2018}
\acmYear{2018}
\acmDOI{XXXXXXX.XXXXXXX}

\acmConference[Conference acronym 'XX]{Make sure to enter the correct
  conference title from your rights confirmation emai}{June 03--05,
  2018}{Woodstock, NY}
\acmBooktitle{Woodstock '18: ACM Symposium on Neural Gaze Detection,
 June 03--05, 2018, Woodstock, NY} 
\acmPrice{15.00}
\acmISBN{978-1-4503-XXXX-X/18/06}

\usepackage{tabularx}
\usepackage{booktabs}
\usepackage{multirow}
 
\usepackage[authormarkup=none]{changes}
\definechangesauthor[name=Lena (Revision 1), color=black]{L1}

\usepackage{xcolor}

\usepackage{layouts}
\hypersetup{bookmarks=false}
\begin{document}


\title{ToMigo: Interpretable Design Concept Graphs for Aligning Generative AI with Creative Intent}

\author{Lena Hegemann}
 \affiliation{%
  \institution{Aalto University}
  \country{Finland}
 }
 \email{lena.hegemann@aalto.fi}

 \author{Xinyi Wen}
 \affiliation{%
  \institution{Aalto University}
  \country{Finland}
 }
 \email{xinyi.wen@aalto.fi}
 
 \author{Michael A. Hedderich}
 \affiliation{%
  \institution{LMU Munich \& Munich Center for Machine Learning}
  \country{Germany}
 }
 \email{hedderich@cis.lmu.de}

 \author{Tarmo Nurmi}
 \affiliation{%
  \institution{Aalto University}
  \country{Finland}
 }
 \email{tarmo.nurmi@aalto.fi}

\author{Hariharan Subramonyam}
 \affiliation{%
  \institution{Stanford University}
  \country{USA}
 }
 \email{harihars@stanford.edu}

\renewcommand{\shortauthors}{Hegemann, et al.}

\begin{abstract}

Generative AI often produces results misaligned with user intentions, for example resolving ambiguous prompts in unexpected ways. Despite existing approaches to clarify intent, a major challenge remains: understanding and influencing AI's interpretation of user intent through simple, direct inputs requiring no expertise or rigid procedures. We present ToMigo, representing intent as design concept graphs: nodes represent choices of purpose, content, or style, while edges link them with interpretable explanations. Applied to graphic design, ToMigo infers intent from reference images and text. We derived a schema of node types and edges from pre-study data, informing a multimodal large language model to generate graphs aligning nodes externally with user intent and internally toward a unified design goal. This structure enables users to explore AI reasoning and directly manipulate the design concept. In our user studies, ToMigo received high alignment ratings and captured most user intentions well. Users reported greater control and found interactive features—editable graphs, reflective chats, concept-design realignment—useful for evolving and realizing their design ideas.
 
\end{abstract}





\begin{teaserfigure}
  \includegraphics[width=\textwidth]{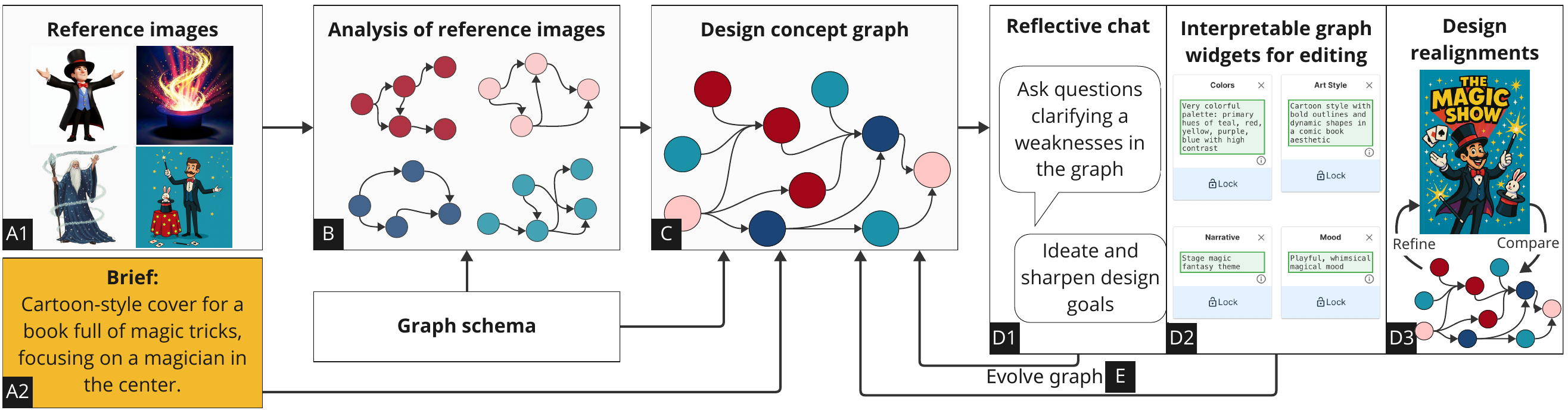}
  \caption{ToMigo represents AI interpretations of user intent as interpretable and controllable design concept graphs. Supporting both reference images (A1) and written briefs (A2), the system analyzes visual features using a structured schema (B) to synthesize a unified design concept graph (C). This graph structure provides a modular framework where individual nodes allow for precise adaptability, while reasoning-based edges ensure internal coherence across the entire design concept. These properties enable three key interaction techniques: surfacing unclear aspects to sharpen intent (D1), providing direct handles for editing AI reasoning (D2), and realigning generated designs with the evolving concept (D3). This workflow allows users to understand and direct the AI's logic as design goals clarify for both the user and the system (E)}
  \Description{The figure shows the components of the ToMigo system as a pipeline from right to left, labeled with letters A, B, C, D, and E. ToMigo uses graphs to make AI’s understanding of user intentions interpretable and controllable in graphic design, supporting reference images and written briefs (A). Box A1 shows an example set of 5 reference images with different comic-style magician illustrations, stage costumes, and prompts such as hats, bunnies, and cards.  Box A2 shows the description ``Cartoon-style book cover of a magician, highlighting a book full of magic tricks, focusing on the magician in the center, in a colorful comic book style.'' under the title ``Brief''. ToMigo analyzes features of each reference image (B) using a graph schema and combines them into a design concept graph (C). Box B shows five graph illustrations in different colors, each representing the analysis of one reference image under the title ``Analysis of reference images''.Box C shows a bigger graph combining nodes of different colors under the title ``Design concept graph''. This graph aligns with user intentions evident in the input and ensures all nodes align with a coherent goal. The graph’s interpretable, adaptable structure enables interaction techniques (D) that help users understand and direct the AI’s reasoning while keeping the graph aligned with intentions as they clarify for both ToMigo and the user (E). E labels two arrows going back from the interaction techniques to the graph, indicating that user information flows back to the graph to update it. The box D is split into three parts labeled D1-3, showcasing one interaction technique each. These include surfacing unclear aspects to prompt alternatives and sharpen intentions (D1 has the title `` Reflective chat'' in a speech bubble underneath it reads ``Ask questions clarifying weaknesses in the graph'' and another speach bubble underneath reads ``Ideate and sharpen design goals''), showing ToMigo’s understanding for direct editing (D2 has a title that reads ``Interpretable graph widgets for editing'' and shows a screenshot of a widget), and effortlessly generating and improving the design as ToMigo realigns it with the evolving graph (D3 shows a magic trick book cover above a concept graph with an arrow pointing from the cover to the graph saying ``compare'' and an arrow pointing back saying ``refine'' and a title ``Design realignments'').} 
  \label{fig:teaser}
\end{teaserfigure}


\maketitle

\section{Introduction}

Controlling generative AI is a known challenge, where the output is often unexpected \cite{10.1145/3563657.3596138,hedderich-etal-2025-whats}. 
In design in particular, intentions are usually not well defined at the start. Designers rely on a design process of refining both the problem and the solution, a process of co-evolution where each informs the other \cite{maher1996modeling, dorst2001creativity}. For example, a designer may begin with a vague goal such as creating a “calm” visual identity, gradually refining this intention (or problem definition) through exploration of color, typography, and layout (solution candidates). This iterative process is associated with better design outcomes but requires navigating multiple co-evolving spaces. Generative AI can disrupt this process by skipping over key steps in developing and communicating design intent.

Three reasons for these challenges have been identified based on the cognitive processes involved in using generative AI \cite{subramonyam2024bridging}:  
(1) Challenges in knowing the exact procedure without taking the actions, which leads to difficulty instructing the AI and getting side-tracked by generated results. This difficulty is obvious for novices, but even experts may struggle to express procedures when tacit knowledge is omitted.  
(2) Challenges with precision when providing the prompt, which lead to unexpected interpretations of the input by the AI.  
(3) Challenges in developing specific intentions from the initial goal, which lead to difficulties evaluating the output.

Most interactive approaches to align generative AI assist users in a translation process. Prompt engineering tools \cite{brade2023promptify, wang2024promptcharm, peng2024designprompt} help users translate their intent into prompts, e.g., helping them discover how their prompt affects the outcome, which properties to target, or how to formulate the prompt. On the other end of the spectrum are tools that enable expression of intent in a more familiar or direct modality (e.g., sketching \cite{Lin2025inkspire}). With ToMigo, we aim at an intermediate approach by explicitly modeling an AI interpretation of the user's creative intent, similar to a theory of mind (ToM), the ability to make and update assumptions about another person’s intentions \cite{goldman2012theory}. The goal is to keep prompting modalities accessible to non-experts of design or image generation and expand under specified prompts coherently.

Building on this goal, a key challenge is that even when prompts are coherent, important aspects of the intended outcome often remain implicit. ToMigo’s approach, grounded in the theory of mind concept, is motivated by the way humans reason about others’ goals using available information. Inspired by this idea, large language models could be prompted to generate intermediate reasoning steps, a process commonly referred to as chain of thought \cite{wei2023chainofthoughtpromptingelicitsreasoning}, which could serve as a technical candidate for reasoning about under-specified prompts \cite{chang2024injecting}. Chain-of-thought reasoning is human readable, allowing users to inspect the AI’s assumptions and understand how it interprets the prompt. However, in order to support iterative design, what is needed is a structured representation of intent that can be revised over time if assumptions are incorrect. In design research, such shared representations are often described as boundary objects \cite{mark2007boundary}, artifacts that are interpretable and editable by multiple stakeholders, which in this context would allow both the AI and users with a theory of mind to collaboratively clarify and refine the intended outcome.

We propose ToMigo, which closes this gap with an interpretable and controllable representation of user intent. \added[id=L1]{In a formative study, we gathered a dataset of user intentions for graphic design generation. The analysis identified ten design intention types with complex interdependencies. For instance, a playful mood suggests bright colors; these colors help establish a cartoon style that reinforces the playful mood. To best represent these relationships, we synthesized a graph schema. Based on this schema, ToMigo represents design concepts as graphs.} A design concept graph is a representation of the design to be generated and contains both extracted intent (design decisions as nodes) and reasoning (edges between them). This representation is interpretable: its modularity makes it easy to explore. It is adaptable: nodes and edges can be edited, added, or removed. Edits can be local—changing details without losing control over the rest of the graph—or global—propagated through the graph along the edges to automatically maintain coherence in the design concept. Designs can be generated from the graph, and existing designs can be edited to reflect changes in the graph.

We contribute a computational method that interprets any combination of verbal and visual inputs, treating them as a unit that reflects partly unspoken user intentions. We propose methods for constructing and refining design concept graphs that evolve with user intentions.  
We further propose interaction techniques that use design concept graphs to support interpretability of AI reasoning and provide direct handles for influencing it. We illustrate this with three interaction techniques: theory-of-mind widgets that provide a direct interface to graph nodes for inspection and editing, a chat interface with clarifying questions that target reflection about the interconnectedness of design decisions, and design generation and regeneration in alignment with the updating graph leveraging the modular structure provided by nodes.

We validate our approach in two user studies. The first study quantifies alignment with initial user goals as a foundation for interactive refinement, while the second explores the effect on intent alignment through a creative process with iteration. Our contributions are as follows:

\begin{itemize}
    \item A graph schema for analyzing, defining, and evolving design concepts that align with user intentions  
    \item A computational approach to interpreting intentionality in sets of reference images and short written briefs, and capturing them in internally coherent design concept graphs  
    \item Three interaction techniques that demonstrate the utility of design concept graphs for aligning generative AI with evolving design intentions  
    \item Findings from two user studies showing that interacting with interpretable design concept graphs supports user control while evolving design concepts  
\end{itemize}
\section{Related Work}

Our work aims to align generative AI with creative intent through a graph-based representation of inferred intent. Here, we review what has been explored before to align AI with creative intent and how design has been represented in graph-based formats. Beyond textual prompts, our work uses reference images as modality to communicate creative intent and we include a review how this modality has been used previously in creativity support tools.

\subsection{AI Alignment with Creative Intent}
Given the existing challenges in aligning AI with user intentions, interaction techniques for steering generative AI towards intended outcomes remains an active area of research. Prompt engineering tools provide processes and support for discovering effective prompts. Promptify \cite{brade2023promptify} supports users in expanding and organizing prompts and generated images. PromptCharme \cite{wang2024promptcharm} and GenTune \cite{gentune2025wang} aid users in connecting which parts of a prompt affected which part of the generated image, thus enabling controlled changes. 

PromptNavi \cite{promptnavi2025huang} analyses the attributes of generated images and supports users in exploring and specifying alternatives. Similarly ContextCam \cite{contextcam2024fan} collects relevant contextual information and proposes expansions before generating images and engaging users in refinements. Specifically for the design context with divergent and convergent phases, DesignPrompt \cite{peng2024designprompt} offers search functionality to discover inspiration and controlling specific details in image generation. Selenite \cite{selenite2024liu}, Lumiate \cite{10.1145/3613904.3642400}, and DesignWeaver \cite{tao2025designweaver} support users in exploring design spaces. Closely related to the theory of mind widgets in our system, is Gmeiner et. al.'s interaction technique around intent tags \cite{Gmeiner2025IntentTEA}, which users can interact with to discover and define creative intentions in slide design. While this and some of the other tools, support users with suggestions for additional intent specifications, they are distinct from the structured theory of mind model of ToMigo which aims at capturing a coherent and editable design concept with reasoning.

Particularly for image generation and generative editing, many tools have been developed to make interactions more intuitive for designers by adapting them to visual modalities. Notable streams of work include sketch-based methods \cite{worldsmith2023dang, Koley2024ItsAAA, Lin2025inkspire}, which increase user control over the layout and brush-based interactions \cite{peng2025fusain, Riche2025aiinstruments, young2023promptpaint} which enable localized AI generated edits to generated visuals. In addition to spacial control, Brickify \cite{shi2025brickify} allows users to reuse design elements by composing them on a canvas. These techniques are particularly useful for controlling the placement or layout of generated image content or style. However, one of the key advantages of text-to-image and image-to-image prompting is the simplicity of the input, which makes it highly accessible to a wide audience. Our work aims to preserve this simple mode of interaction by adding a theory-of-mind representation of user intention, making it interpretable and manipulable. 

\subsection{Graph Representations of Design}

Our method uses graphs to represent design concepts, which is inspired by multifaceted ways in which prior work has used graph structures to represent designs and design processes for computation and visualization. GUIs can be represented as trees where nodes and child nodes indicate how design elements are nested in each other especially for mobile apps \cite{li2020autocompletionuserinterface, squire2025leung} and websites \cite{dom2002marini} which helps computationally modifying them. To further aid modifications of layouts, Jiang et al. \cite{jiang2024graph4gui} trained a graph neural network to predict constraints such as alignment or grouping in GUIs. Graph-based representations have also been used for setting constraints for design attributes such as color palettes \cite{Mellado2017ConstrainedPEA} to enable automatic editing. 
Entity relationship diagrams are a form of graph used to describe interrelated data with nodes (entities) and edges (relationships). In the context of generative AI, these have been used to visualize structures in generated text \cite{jiang2023graphologue} and stories \cite{toward2025qin, xcreation2023yan} making it explorable and globally editable. Graphs as a visualization of structure have also proven useful to show the evolution of design ideas \cite{coexploreds2025chen, ideationweb2025shen}, or for interactive versioning and remixing designs \cite{worldsmith2023dang, angert2023spellburst, promptnavi2025huang, mixplorer2022kim}. 

In StoryEnsemble \cite{storyensemble2025suh} the outcomes of different concept design stages, such as personas, problem and solution definitions, are represented as nodes and linked to each other so information can propagate forward and backward, supporting iteration. Kim et al. \cite{kim2025plantogether} developed an information graph to support AI application development by defining information and dependencies to support designers in defining a complete AI application plan. ToMigo combines the idea of an information graph as a schema with that of propagating changes through notes. However, our schema is also used as a framework for intent analysis, and the edges in our design concept graphs are augmented with interpretable reasoning.

\subsection{Tools for Designing with Reference Images}
Designing with reference material such as inspirational images is a common practice, and many tools have been developed that build on this technique. Some are specifically designed to help users find and collage reference material as an ideation method \cite{koch2019may}, enhance reference material to explore a design vision \cite{koch2020semanticcollage, imagesense2020koch}, dimensions of the design space \cite{designfromx2025duan, fashionq2021jeon} and to communicate with collaborators during a creative process \cite{chung2023artinter}. Another stream of research explores the use of reference material directly in creation processes such as image generation \cite{shi2025brickify, peng2024designprompt}, UI design \cite{lu2025misty}, illustration \cite{choi2024creativeconnect}, data visualization \cite{dataquilt2020zhang}, songwriting \cite{amuse2025kim} and composing visual styles \cite{zhou2024stylefactory}. While our work shares with these tools the goal of transferring specific elements from references into a design with fine-grained control, our focus is on providing a transparent representation of AI understanding and ensuring alignment with design intent more broadly.

\section{Design Considerations}\label{design-considerations}
\begin{figure*}
    \centering
    \includegraphics[width=\textwidth]{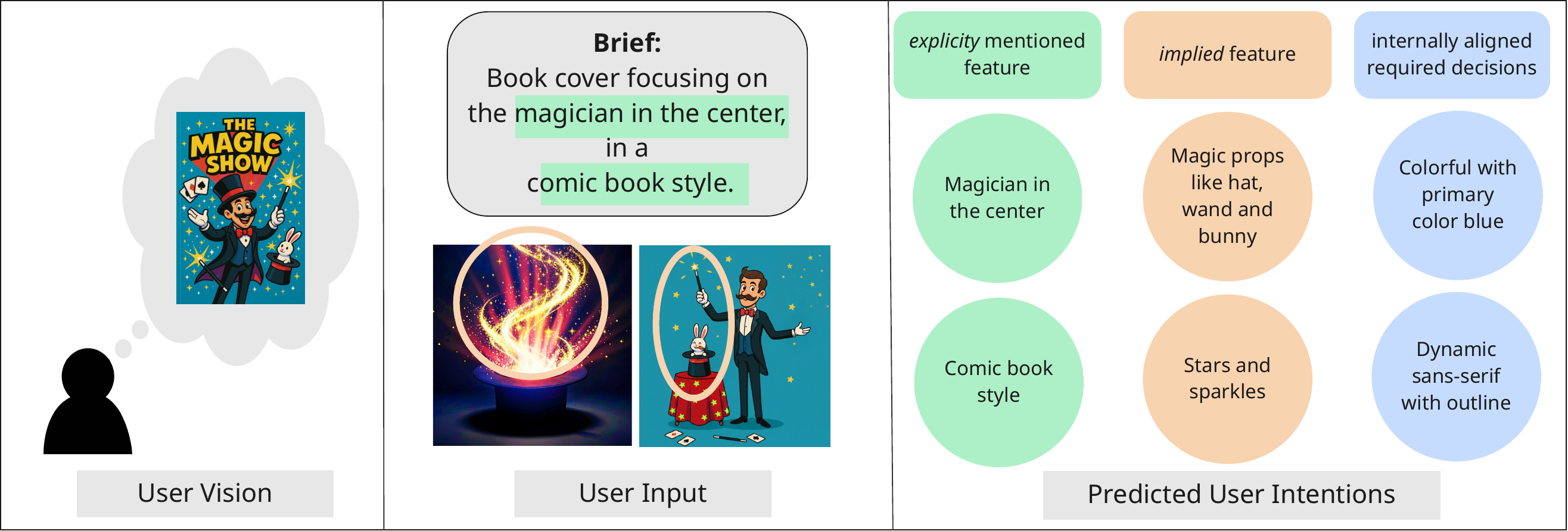}
    \caption{Predicting user intentions correctly is a communication task where a user conveys their vision (left) incompletely with simple text and image input (middle) from which the system needs to reconstruct it. The model of is expected to correctly represent features that are explicitly mentioned but to reconstruct the vision fully, it will need to identify implied features (for example repeated or salient features in the images) and reason about choices for features that are less clearly indicated in the input but required to realize the explicit and implied features.}
    \label{fig:scoping}
    \Description{Diagram of user vision, input, and predicted intentions in the design process.
    On the left, “User Vision” is illustrated with a cartoon-style book cover of a magician labeled The Magic Show. In the middle, “User Input” includes a short design brief—“Book cover focusing on the magician in the center, in a comic book style”—and two example images showing a magician with props. On the right, “Predicted User Intentions” are represented in three categories: explicitly mentioned features (magician in the center, comic book style), implied features (magic props such as hat, wand, bunny, stars and sparkles), and internally aligned required decisions (colorful with primary blue, dynamic sans-serif font with outline).
    }
\end{figure*}

We base ToMigo on the principle of problem-solution co-evolution, assuming that the vision of the user and the output of the AI evolve in tandem. In this process ToMigo aims to support users in developing and communicating specific design intent to achieve their desired outcome from the interaction with AI. The idea is that ToMigo proactively predicts user intent through a theory of mind model which the user can interpret and edit, such that it simultaneously functions as a boundary object shared between user and AI. This approach is specifically aimed to support novice designers who benefit from using simple inputs such as natural language or unedited images but might require a scaffold to articulate complex design decisions. The following design considerations map these theoretical goals to concrete requirements for the theory of mind model. 

\begin{itemize}
    \item[\textbf{DC1}]\textbf{Aligned with user intentions.}
    As the core requirement for the \emph{theory of mind} model, the system must accurately capture the design intent expressed and implied in the user input. Explicit elements within the prompt should appear faithfully in the model while inferred intentions should remain consistent with user expectations. For example, if a user specifies a main motif and a specific mood, the system should reflect both while suggesting stylistic choices that naturally suit that mood. Although perfect inference from under-specified prompts is difficult, the goal is to reason beyond the prompt as reliably as possible. The system should be flexible to maintain accuracy as new information becomes available.

    \item[\textbf{DC2}] \textbf{Ensure coherence by internally aligning design choices.} 
    The system should maintain a coherent and complete design concept by identifying implied elements that harmonize the overall vision. For instance, if a user provides a title for a book cover, the system should reason why specific typographic choices as fit a requested mood or style rather than as isolated requirements. This internal consistency and reasoning should support novice designers by delivering design concepts that follows sound design logic. Furthermore, grounding the generation process in these articulated, coherent decisions should improve the reasoning quality of the AI and lead to more accurate model outputs \cite{wei2023chainofthoughtpromptingelicitsreasoning}.
    
    \item[\textbf{DC3}] \textbf{Provide an interpretable and inspectable representation.} 
    The representation should be structured in a way that makes the components of the design concept explorable as a functional \emph{boundary object}. This shared artifact should allow users to inspect what the system predicted about the design without assuming design or AI expertise. For example, a design concept may expose its selected color palette, composition, and function as distinct but related data points. This form of interpretability should provide \emph{novice support} by presenting a design structure and the domain-specific vocabulary that helps users to articulate their design intentions. By making the computational processing of input transparent, the representation should encourage users to reflect on their goals and provide the handles for iterative refinement.
    
    \item[\textbf{DC4}] \textbf{Support modular adaptability and refinement.} 
    The system should maintain a modular structure to accommodate the \emph{evolution} of a design concept as user intentions shift over time. This flexibility should allow fine-grained adjustments such as changing, adding, or removing individual parts of the concept. When these changes occur, the system should assess documented connections to determine if additional updates or user grounding regarding realignment become necessary. For example, changing an art style from comic to photorealistic would likely affect textures or color schemes, while modifying a character posture might not require further changes to other decisions.
    
    \item[\textbf{DC5}] \textbf{Enable iterative and grounded generation.} 
    The system should use its internal model to generate aligned designs and support continuous refinement through targeted updates. To facilitate \emph{co-evolution}, the system must react to the evolved user intent by realigning the design with the updated \emph{theory of mind}. In this process, the representation should serve as a \emph{boundary object} that the system interprets to determine how the current design state differs from the theorized intent. For example, if a user interaction changes the desired text or composition in the \emph{theory of mind} model, the system should identify the discrepancy to the existing design. By comparing the model to the current state of the design, the system should realize the necessary changes to bring the two back into alignment.

\end{itemize} 
\section{Formative Study}\label{schema}
This section describes our formative study to develop the structure of the design concept graph. We collected a dataset of user intentions to construct a graph schema that supports novice graphic designers who work with verbal briefs and reference images. 

Novice designers rely heavily on external resources, often exceeding the usage rates of expert designers \cite{gonccalves2011around}. However, prior work suggests that novices reflect less on these resources and instead adopt inspirational material more directly \cite{garner2001problem}. Despite these observations, we know little about which specific features novices prioritize in inspirational images or how those features shape their design intentions. Our formative study identifies the features novices seek and captures them in a structured representation to serve as a schema for design concept graphs.

We gathered insight into how users use reference images by collecting a dataset of images paired with verbal ground truth. This study addresses the following research questions:
\begin{enumerate}
    \item[\textbf{RQ1:}] Which specific features do users intend to transfer from a reference image to a new design?
    \item[\textbf{RQ2:}] How do users refer to these features using a combination of visual and linguistic cues?
    \item[\textbf{RQ3:}] How do users perceive the underlying relationships between these different design features?
\end{enumerate}

\subsection{Data Collection}
We gathered 100 responses from US-based, English-speaking crowd workers via Prolific\footnote{https://www.prolific.com/ accessed on Jan 19, 2026}. Each participant received 1.50 USD for the survey, which averaged 7 minutes to complete.

The survey prompted participants to imagine commissioning a graphic designer for a birthday invitation. This topic provided a familiar context while allowing for a wide variety of styles and themes. Our instructions required participants to upload three high-quality, distinct images to serve as inspiration for the designer. We asked them to consider various design aspects in the images and specifically explain for each image how a designer should it. These instructions aimed to elicit detailed explanations that would provide actionable information.

We reviewed the data quality of every response and excluded one sample due to poor image quality. The final dataset contains 99 sets of three image-intention pairs. Because participants provided their inspiration in an open-ended format, the response lengths varied. These descriptions ranged from single words, such as ``colors,'' to full paragraphs detailing multiple features and specific design reasoning.

\subsection{Graph Schema Construction}

We constructed an information graph in two rounds using principles of thematic analysis and affinity diagramming \cite{lucero2015using}. During the first round, we reviewed each image-explanation pair to identify the inspirational features it contained (RQ1). We recorded these features as nodes and used the participants' original wording for the examples and many of the codes (RQ2). We organized the nodes on a virtual canvas, grouping similar concepts. 
If we noticed that different explanations used the same wording to mean different things, we split the nodes and defined our own codes while keeping the user's language in the examples (e.g. brightness referring to color chroma in one sample and to lightness in another resulted in two nodes \emph{chroma, e.g. brightness} and \emph{lightness, e.g. brightness}). This strategy increased precision while we continued to track user terminology. For repeated features, we tracked their frequency and added new examples to the existing nodes.

We also mapped how users perceive the relationships between these features (RQ3). When an image-explanation pair indicated a connection between concepts, we drew a line between the nodes and attached descriptive codes to the line using the same wording strategy. We analyzed 50 sets (150 image-intention pairs) until we observed saturation, leaving the remaining 49 sets for tuning the system during development. 

In the second round, we reviewed the groupings, defined distinct concepts, and introduced supercategories for the grouped nodes. A few codes were split or merged in this iteration if the examples indicated that the word vaguely referred to multiple concepts (e.g. \emph{`look'}) or had only few mentions while fitting well into another category (e.g. \emph{`info'} was merged with \emph{`verbal content'}). 

We also unified the direction of the edges to express \emph{supportive influence}. While we chose to represent how low level features support higher level goals, these relationships also function as requirements that flow in the opposite direction. For example, an edge showing that a specific color palette supports a certain mood can be reformulated as a requirement where that mood calls for a specific palette. We unified the direction in the schema to avoid duplicate edges that describe the same relationship from different perspectives. This design decision simplifies the graph structure and prevents maintenance difficulties that arise from redundant connections. We chose the upward, supportive orientation because it aligns with how our participants most frequently described their design reasoning. 

We transcribed the graph into a text format that includes specific definitions for every code. This document appears in the Supplementary Material. Because the graph guides how an LLM reasons over user intentions, we used LLM feedback to refine our definitions. We prompted both Claude and GPT to explain the relationships between nodes to verify they used the definitions accurately. We iterated on category names and phrasing until the models consistently applied the logic of the graph. After finalizing the text based on LLM performance, three human evaluators reviewed the definitions to ensure they remained distinct and clear.

\subsection{Graph Schema}

\begin{figure*}
    \centering
    \includegraphics[width=0.8\textwidth]{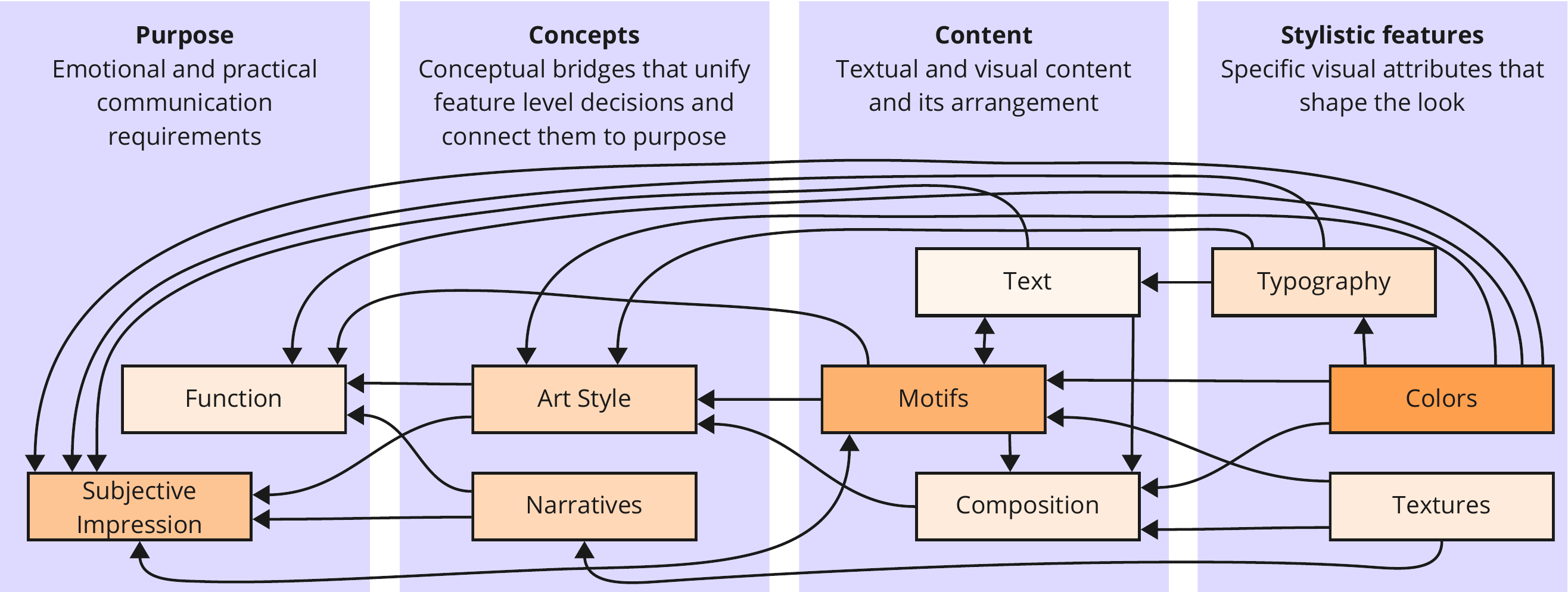}
    \caption{The graph schema organizes node types (orange) into four primary roles (purple). Lines indicate relationships found in our dataset where participants mentioned aspects in relation to each other. More saturated orange indicates a higher frequency in the dataset. Connections typically flow from right to left as low level feature decisions support choices that concern the design as a whole.}
    \label{fig:graph-overview}
    \Description{A diagram shows the four role nodes can have in the graph schema and in each shows the specific node types. On the far right, Stylistic Features include color, typography, and textures. Moving left, Content nodes cover motifs, composition, and verbal elements. These lead into Concepts nodes, which include art style and narratives. On the far left, Purpose nodes represent functional goals and subjective impressions. Arrows indicate the flow of influence from granular features toward holistic design goals.}
\end{figure*}

The final graph schema organizes design concepts into nodes and defines their relationships through directed edges. We arrange these nodes along a spectrum that ranges from granular feature-level decisions to holistic design goals. This hierarchy allows the system to bridge the gap between specific visual choices and the broader communicative purpose of a design.

\subsubsection{Nodes}
We categorize the nodes into four primary roles as illustrated in Figure \ref{fig:graph-overview}: \emph{Purpose}, \emph{Concepts}, \emph{Content}, and \emph{Stylistic Features}. \emph{Purpose} nodes capture high-level goals by defining the intended function and the emotional impression of the design. \emph{Concepts} nodes provide a unifying framework through specific art styles or narratives. \emph{Content} nodes describe the physical arrangement of the composition, including motifs and layout. Finally, \emph{Stylistic Features} nodes define granular visual attributes like color palettes and typography. Table \ref{tab:node_schema} details each node type and defines its specific function within the schema.

\begin{table*}[ht]
\centering
\caption{Node Types in the Schema Organized by Role}
\label{tab:node_schema}
\small
\renewcommand{\arraystretch}{1.5} 
\begin{tabularx}{\textwidth}{l l X}
\toprule
\textbf{Role} & \textbf{Node Type} & \textbf{Description} \\
\midrule
\multirow{4}{*}{\shortstack[l]{\textbf{Purpose}}} 
& Subjective impression & Relates to how the design is interpreted subjectively. These could be moods which describe an emotional response, ambiance related to a social or cultural setting evoked, or character expressed by the design. \\ \addlinespace
& Function & Refers to the purpose or intended outcome of the design, focusing on how it communicates, appeals to specific audiences, and serves practical or thematic goals. \\
\midrule
\multirow{4}{*}{\shortstack[l]{\textbf{Concepts}}} 
& Art Style & A consistent artistic style or aesthetic approach that shapes how the visual elements are interpreted and combined. \\ \addlinespace
& Narratives & Broader concepts, themes, or genres that group a design’s underlying ideas, narratives, and emotional tones. They provide a conceptual framework that guides the visual and contextual elements. \\
\midrule
\multirow{6}{*}{\shortstack[l]{\textbf{Content}}} 
& Motifs & Visual elements within a design that serve as objects, symbols, or shapes to convey meaning, enhance the narrative, or provide decorative details. Subcategories include the main subject and its expression, decorative objects, and structures such as borders. \\ \addlinespace
& Composition & Positioning and sizing of the visual elements, including foreground, middle-ground, and background. \\ \addlinespace
& Verbal elements & Refers to the verbal contents and contains considerations of what information is displayed in titles or other texts, as well as diction and verbal tone. \\
\midrule
\multirow{6}{*}{\shortstack[l]{\textbf{Stylistic}}} 
& Colors & The palette of colors used in the design, with subcategories describing specific inspirational features within the palette (colorfulness, contrasts, lightness, chroma) or specific application areas (primary colors, background colors, object colors). \\ \addlinespace
& Typography & The visual and aesthetic characteristics of text in terms of font families and styling, custom lettering, or hand-drawn styles. \\ \addlinespace
& Textures & Surface qualities of motifs and backgrounds. \\
\bottomrule
\end{tabularx}
\end{table*}

\subsubsection{Edges} Edges represent the relationships between inspirational aspects and carry specific labels to describe how these elements interact. These connections explain how one design choice influences another, such as when a specific color palette ensures the readability of typography or a motif acts as a symbol within a narrative. Our data shows that most relationships flow from stylistic features and content toward conceptual frames and final purposes. For example, participants frequently noted that artistic style impacts both content and purpose, while content choices directly influence the overall purpose. We also observed that feature level types impact conceptual frames, such as colors influencing typography or composition affecting the arrangement of motifs. Although these relationships can function in both directions, we unified the edges to follow this upward flow because it represents the most frequent pattern in our dataset. This structure provides an overview of how low level visual decisions typically support the broader goals of a design.
\begin{figure*}
    \centering
    \includegraphics[width=\linewidth]{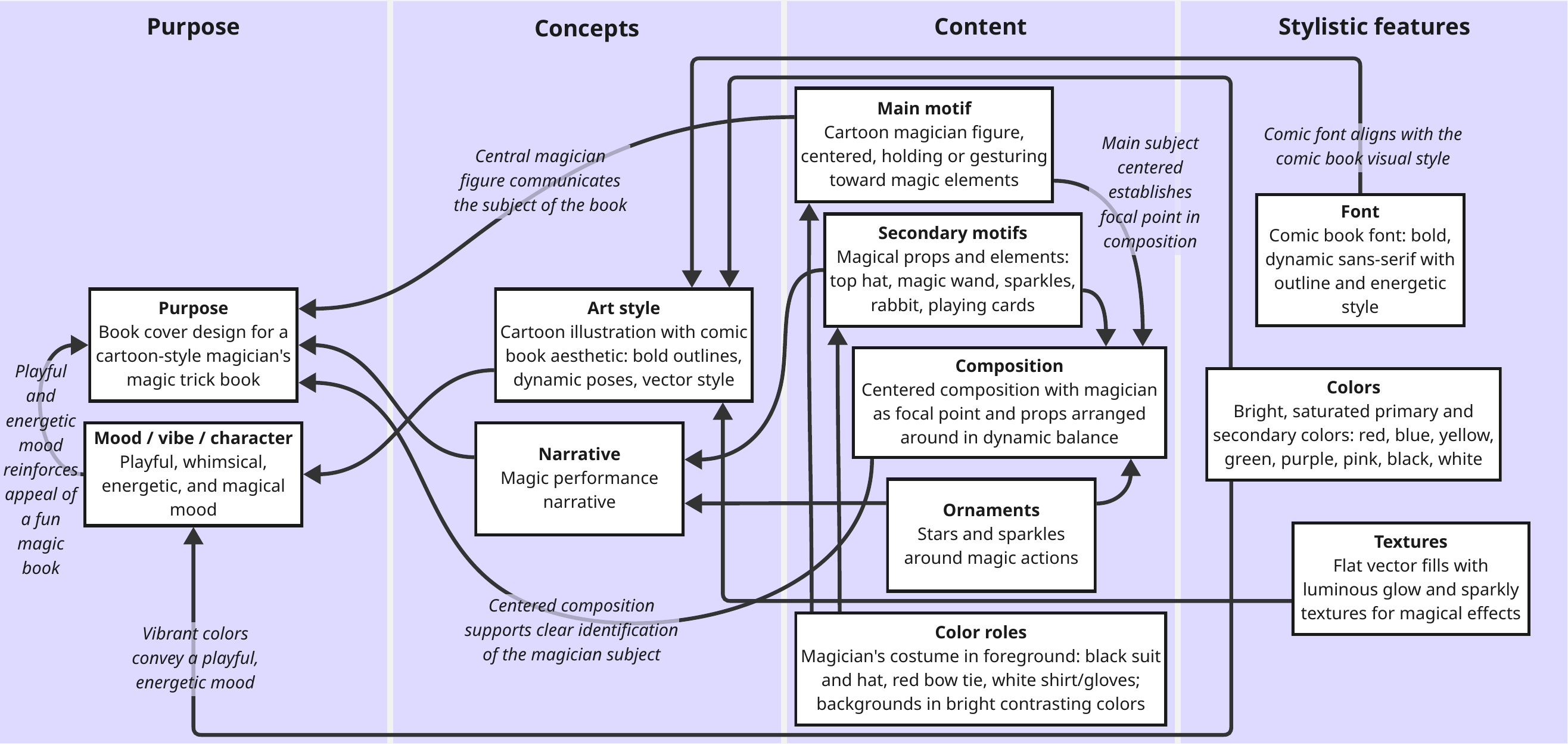}
    \caption{Example of a design concept graph. This concept graph instantiates the relevant node types with concrete descriptions of design decisions for the magician book cover. Edges are labeled with reasons why the source node supports the target node. All edges have reason labels. Due to space constraints the figure does not show all of them.}
    \label{fig:graph-representations}
    \Description{Design concept graph for a magician-themed book cover.
    The graph is organized into four columns: purpose, concepts, content, and stylistic features.
    -	Under Purpose, nodes describe the design as a cartoon-style magician’s magic trick book with a playful, whimsical, energetic, and magical mood.
    -	Under Concepts, nodes include art style (comic book illustration with bold outlines and vector style) and narrative (magic performance).
    -	Under Content, nodes describe the main motif (magician figure centered with magic elements), secondary motifs (props such as top hat, wand, rabbit, playing cards), composition (centered magician with props in dynamic arrangement), ornaments (stars and sparkles), and color roles (contrasting costume and background colors).
    -	Under Stylistic features, nodes include comic-style font (bold, dynamic sans-serif with outline), bright saturated colors (red, blue, yellow, green, purple, pink, black, white), and flat vector textures with sparkly effects.
    
    Edges connect nodes across columns with labeled reasons, such as how centered composition supports clear identification of the magician or how vibrant colors reinforce a playful mood.
    }
\end{figure*}

\section{Graph Representations of Designs}

To illustrate system's functions, this section provides a consistent walkthrough using a specific design task as a running example inspired by a user in study 1 (Section \ref{study1}. In this example, the user wanted to create a cartoon-style book cover for a collection of magic tricks and provided a short design brief and reference images of a magicians with various props, some of them in cartoon style (see Figure \ref{fig:teaser}). ToMigo then processes these inputs to build a structured representation of the project. The following subsections describe how the system transforms this initial input into a detailed concept graph and uses that graph to guide the interaction and generation of the final book cover.

ToMigo represents design concepts as graphs based on the graph schema. The system populates this structure by describing specific properties of a design as individual nodes. A node is defined by a unique ID, the category label (as in Table \ref{tab:node_schema}) and a textural description. For example for the magician book cover, the system identifies the "Cartoon-style" request and creates a corresponding node like \texttt{ID: 1, Label: Art Style, Description: Cartoon illustration with comic book aesthetic bold outlines dynamic poses, vector style}. Edges connect these nodes to provide a rationale for how different design decisions harmonize within the same project. They link to the node IDs and have a rationale. For example, if the system includes a vibrant primary blue for the background, an edge to the cartoon-style node could explain that the primary color is typical for the style and thus emphasizes it. 

The use of the schema remains flexible. Not all designs are expected to contain instances of all nodes. For example, some designs might be typographic compositions without motifs while others have no text. On the other hand, it is possible to extend the graph with custom node categories. Flexibility is also applied to the use of edges. While the schema contains the edges that we observed in the dataset, they are interpreted as frequent/typical edges that are a starting point but not exhaustive.

ToMigo operates on these graph representations of designs to maintain a theory of the user intentions (theory of mind) during design conceptualization. An example can be seen in Figure \ref{fig:graph-representations}. All prompts are provided in the supplementary materials. 

\subsection{Initialization of Design Concept Graphs}
When presented with a new design task, ToMigo initializes a design concept graph based on the provided context (text and image input). The procedure is designed assuming intentionality behind all provided input, thus everything provided by the user is ensured to impact the graph.

\subsubsection{Image Analysis:} If there are images, a multi-modal LLM performs an analysis of the images first, creating one graph per image. Besides the images, it reads the definitions in the schema and four example graphs, and is prompted to create a node for all categories that are present in the image and edges describing how they work together. The result is a concept graph for each image: a list of nodes that describe the image in a structured format, with labels, descriptions and IDs, and edges that link them together referencing the IDs and providing a reason of how the nodes support each other in the image. In the magician example there would be five such graphs.

\subsubsection{Concept Graph Synthesis:} In the next step, ToMigo combines the images with the textual input, ensuring that everything explicitly mentioned will be represented by a node and that each image impacts the concept graph with at least one feature. Our formative study showed that different users have different approaches to providing images ranging from combining a number of example designs of the same type, for example three birthday invitation cards that they like. Some selected a set of images focusing mostly on one of mood, content or style emphasizing what they intended through repetition, for example three different images with balloons. Others introduced specific ideas through images that prominently displayed them for example among a set of photos added a typography sample. ToMigo builds on this observation of images provided in sets: intent of features can be implied by repetition in multiple or saliency in one image. The magician example repeats magic props and sparkles in multiple images which ToMigo identified as secondary motifs. 

The prompt contained the image concept graphs, the images, and instructions for picking features according to these observations and the prompt, and connecting them with new reasoning edges. To avoid the LLM from hallucinating unintended features, we prompted it to quote evidence for their inclusion with reference to the images and textual input. 

\subsection{Theory of Mind Updates}
Given new information, such as a follow up prompt, ToMigo updates the graph. For example the user might say that they wish for more sparkles on the magician book cover. Taking the new input and the history of earlier input, ToMigo is instructed to interpret the intent behind the new input and identify which nodes to add or edit to reflect the intention. To avoid cluttered graphs, priority was given to editing nodes. In this example, edit the node for secondary motifs over adding a second node of this type for more sparkles. Finally, the LLM was instructed to check consistency in the graph and update related nodes to fix inconsistencies. In case of the sparkles, all other nodes would still be aligned. 
\section{User Study 1}\label{study1}
\begin{figure*}
    \centering
    \includegraphics[width=\textwidth]{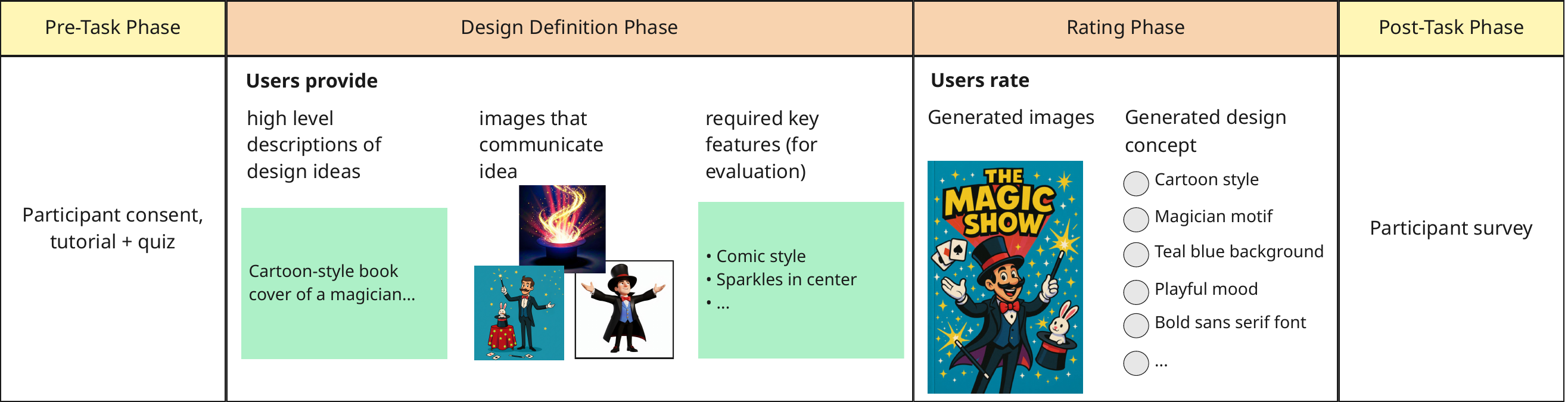}
    \caption{The study procedure consisted of four phases: The pre-task, design definition, rating, and post-task phases. The design phase and the rating phase were repeated three times per participant, resulting in three rated design ideas per participant.}
    \label{fig:study1-procedure}
    \Description{Diagram of the study procedure across four phases.
    The pre-task phase included participant consent, a tutorial, and a quiz. In the design definition phase, participants provided high-level descriptions of design ideas, uploaded example images, and specified required key features for evaluation. The rating phase involved evaluation of generated designs and design concept graphs, with generated images and graph nodes shown. Finally, the post-task phase consisted of a participant survey. 
    }
\end{figure*}

\begin{figure*}
    \centering
    \includegraphics[width=\textwidth]{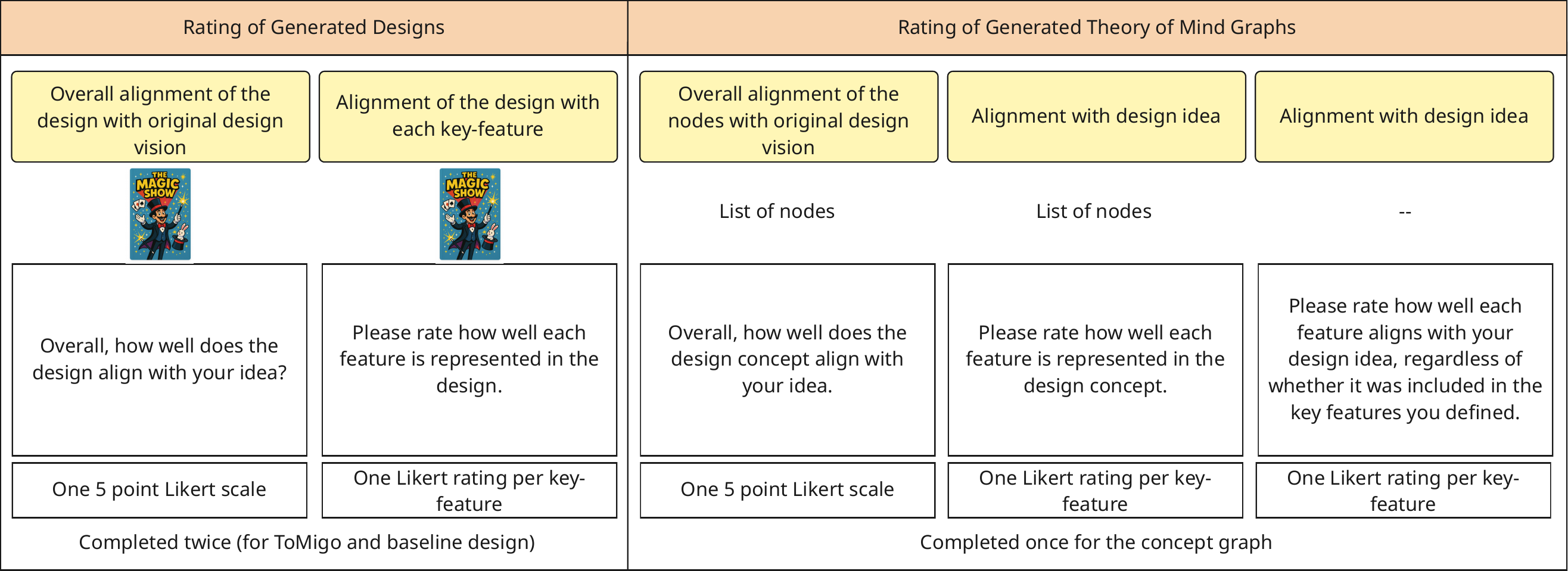}
    \caption{In the rating phase, participants first rated the generated designs for alignment overall and per key-feature defined by them in the design definition phase. After completing these ratings for both designs, participants proceeded to rate the design concept graphs. Given the nodes as a list, as for the designs they rated the graph's overall and per-feature alignment with the idea. Finally, they rated the alignment of each node with their idea. }
    \label{fig:rating-phase}
    \Description{Diagram of the rating phase for designs and concept graphs.
        The figure is split into two sections. On the left, “Rating of Generated Designs” includes two tasks: (1) rating the overall alignment of a generated design with the participant’s original design vision, and (2) rating how well each key feature is represented in the design. Both tasks use 5-point Likert scales and are completed twice, once for the Baseline design and once for the ToMigo design. On the right, “Rating of Generated Design Concept Graphs” includes three tasks: (1) rating the overall alignment of the set of nodes with the original design vision, (2) rating how well each feature is represented in the design concept, and (3) rating how well each node aligns with the design idea, regardless of whether it was part of the original key-features. These graph ratings are also completed using 5-point Likert scales.
    }
\end{figure*}
We conducted a user study to evaluate how closely the concept graphs generated by ToMigo align with user intentions. Since the initial prediction of the user intentions at the start of the design process sets the foundation for the interaction and tells about the system's ability to extract intentions from user input of verbal and visual form, this study is focused on quantifying the match between early project user intentions and the initial design concept graph. While the alignment of the graph itself is our main goal as a foundation for transparent AI understanding of user intent enabling interactions, we further want to evaluate the effect of the graph on the alignment of the generated image artifact itself. Image generation is sensitive to the structure, phrasing, and length of the prompt. In particular, for longer prompt lengths, image generation models might struggle carrying out all instructions~\cite{brade2023promptify,10262331}. We want to ensure that the graph with all its details reaches at least equivalent alignment as a strong baseline with state-of-the-art prompt augmentation.

The following research questions guided the design of the study:

\begin{itemize}
    \item[RQ1:] How well is the design concept graph aligned with user intentions?
    \item[RQ2:] How does the design concept graph impact the alignment between user intentions and generated designs?
\end{itemize}

\subsection{Participants}
We recruited internet users fluent in English via a crowd sourcing platform\footnote{https://www.prolific.com/, accessed on July 2, 2025} sampling globally. To ensure diversity in the sample, we paused the study every 20 participants and, for the remaining participants, excluded regions that started to be overrepresented based on their population size. A quiz was implemented about the task instruction to ensure participants understood what was asked from them. Participants who failed the quiz a second time after receiving feedback on their first attempt, were not permitted to proceed to the main task of the study. See Section \ref{study1procedure} for details about the checks and instructions. In total, we collected 120 responses that completed the main task. We compensated each participant with GBP 12 for a time investment of 80 minutes.

\subsection{Materials}
The study was conducted using a web interface that walked participants through a tutorial and quiz, and allowed them to define design ideas, upload reference images, describe intentions in terms of key features, rate predicted intentions and generated designs, and a post-task survey with three multiple-choice questions about their experience and interest in AI and design. Attention checks were included in the rating phase of the study.

We aimed to collect a dataset of diverse graphic design intentions that can be used to assess the generalizability of ToMigo within the broad field of graphic design. Thus, we developed a set of 18 graphic design tasks from which three were randomly sampled for each participant. Each task specified a different type of design (e.g., poster, logo, package) and matching themes (e.g., festival, personal project, everyday items) to ensure broad coverage of graphic design types and application areas. Instead of prescribing specific topics in narrow briefs, which could be difficult to relate to for some users in our general sample, the tasks provided themes to give participants a direction but allow them to pick their own topic (e.g. for the theme ``festival'', users defined the topic of the festival). For each design idea submitted by participants, a concept graph and two designs were generated, one with ToMigo and one baseline design.

\subsection{Design}
The study used a within-subjects design, where each participant rated two designs, one generated using the ToMigo model and one baseline design, for every design task. For the baseline design, we removed the step of generating a concept graph before generating the design. Instead of the graph, it used the project description directly to generate the design. Both conditions used the same image generator (GPT 4.1 with GPT image tools). The GPT API applied a prompt optimizer in both cases to rephrase the input; for underspecified prompts, it additionally expanded the content.

\subsection{Procedure}\label{study1procedure}

The study procedure consisted of four phases -- an introduction phase, a design definition phase, a rating phase, and a post-task phase (as visualized in Figure~\ref{fig:study1-procedure}).

In the introduction phase, participants were informed about the purpose and structure of the study. To ensure that participants understand how to complete the task, this was followed by a tutorial explaining the required steps for the main tasks. After the tutorial, participants completed a quiz to assess their comprehension of what was asked from them. The quiz contained practice tasks and two multi-select questions. When participants made incorrect selections on a multi-select question, they were provided with the correct answers and a second attempt. Participants failing this attempt were requested to depart from the study.

In the design definition phase, participants defined three design ideas taking three steps for each. First, they were given a randomly sampled design task and asked to specify the type of design and give a high-level description (brief) of their design idea. Second, they were asked to search for inspirational images online and upload a selection of four to six images that communicate their design idea visually. Third, participants were asked to imagine the design and how it is inspired by the images before providing a set of four or more key features they intended for the design to have. They were informed that they will use key features to evaluate the design and instructed to formulate them in such a way that they were mutually exclusive, not in conflict with one another, and specific enough to evaluate whether they are present in a generated design.

Whenever a participant completed a design definition, the design type, brief and images were used to generate a concept graph, a design based on the concept graph (ToMigo design), and a baseline design. The key features were not involved in the generation but reserved for evaluation.

After defining three design ideas, participants proceeded to the rating phase, in which participants rated the alignment of the generated designs and the concept graph (see Figure~\ref{fig:rating-phase}). They rated the two designs first, followed by the graph. The order of rating the designs was randomized and the participants were not aware of the different generation processes. For each design, they first rated how well the design aligned with their original vision (overall alignment). After this, they were shown their key features alongside the image and asked to rate for each feature on how well it is represented in the design. Then, they were shown a list of the node descriptions from the concept graph for which they were asked to provide the same two ratings. Finally, they rated the alignment of each node with their design idea (regardless of whether it was included in their previously defined key features or not).

In the last phase, participants filled out the post-task survey.

\subsection{Data Collection}
The following data was collected from participants
\begin{itemize}
    \item Design ideas: design type, short brief, and reference images.
    \item Key features: four or more features that participants intended for the design, specified after providing the brief and images but before generation.
    \item Design alignment ratings (RQ2): for each design, participants rated overall alignment and alignment with each key feature on a 5-point Likert scale.
    \item Concept alignment ratings (RQ1): for each node in the concept list, participants rated overall alignment as well as alignment with each key feature and with the node itself.  
    \item Prior relevant experience: three ratings of the participants' interest in design, and prior experience with text-to-image generation and design.
\end{itemize}

\subsection{Analysis}
Before analysis, we checked the quality of the responses. Two responses were removed because they failed more than one attention check. In addition, we tested the authenticity of free text responses using an AI detector\footnote{https://gowinston.ai/, accessed on August 28, 2025} and removed 17 submissions with a high AI score. Two submissions were excluded from analysis due to nonsensical key features. Some design ideas were rejected by GPT content moderation. This affected 17 design ideas for which no designs could be generated and rated by the user. In total, we included 280 valid design ideas with ratings in the analysis.

We compared the alignment ratings for the ToMigo and baseline designs. To account for varying amounts of key features defined by the participants, we averaged their ratings per design before comparing the ratings of the two designs. As each participant contributed multiple design ideas across the 18 tasks, we accounted for their potential correlation by choosing mixed effects models in the analysis.

\subsection{Results}
\begin{figure*}
    \centering
    \includegraphics[width=\textwidth]{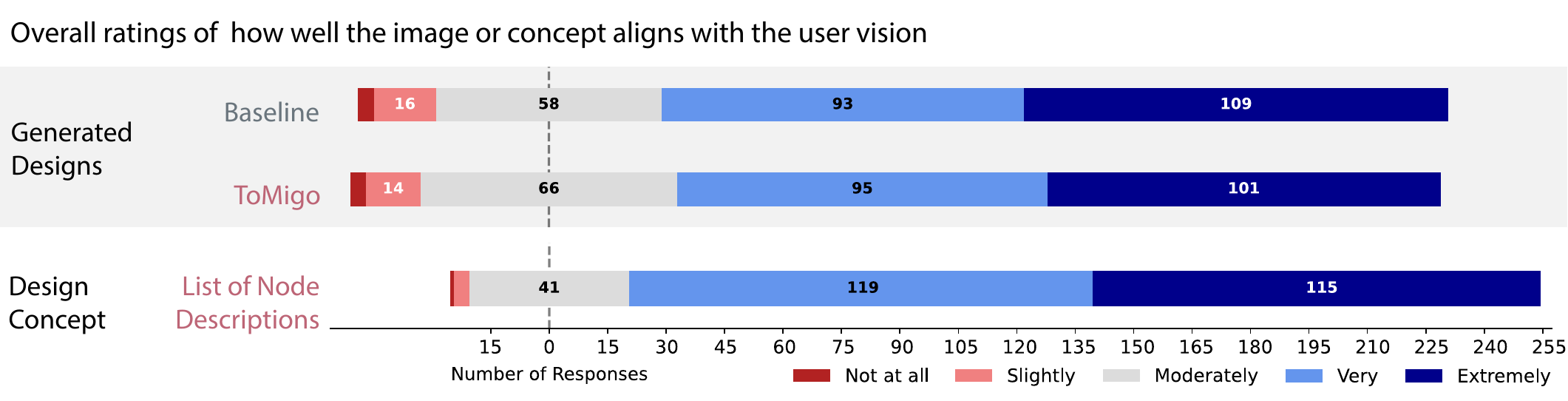}
    \caption{User ratings of how well the output is aligned with their original vision overall. Generally, most designs and the concept graphs were rated very or extremely aligned. The ratings of the two designs received similar ratings (top).}
    \label{fig:overall-ratings}
    \Description{Stacked bar charts of overall user ratings for generated designs and the concept graph. 
                Top chart (Overall Ratings of Generated Designs): 
                For the Baseline design, most responses indicated strong alignment: 93 “very” and 109 “extremely.” Moderate ratings numbered 58, while lower alignment was rare, with 16 “slightly” and very few “not at all.”
                The ToMigo design showed a nearly identical pattern, with 95 “very,” 101 “extremely,” and 66 “moderately,” plus 14 “slightly.” 
                Bottom chart (Overall Ratings of the Concept Graph):
                The concept graph was rated highest overall, with 119 “very” and 115 “extremely,” and only 41 “moderately.” Ratings of “slightly” or “not at all” were minimal across all conditions.}
\end{figure*}

\begin{figure}
    \centering
    \includegraphics[width=\linewidth]{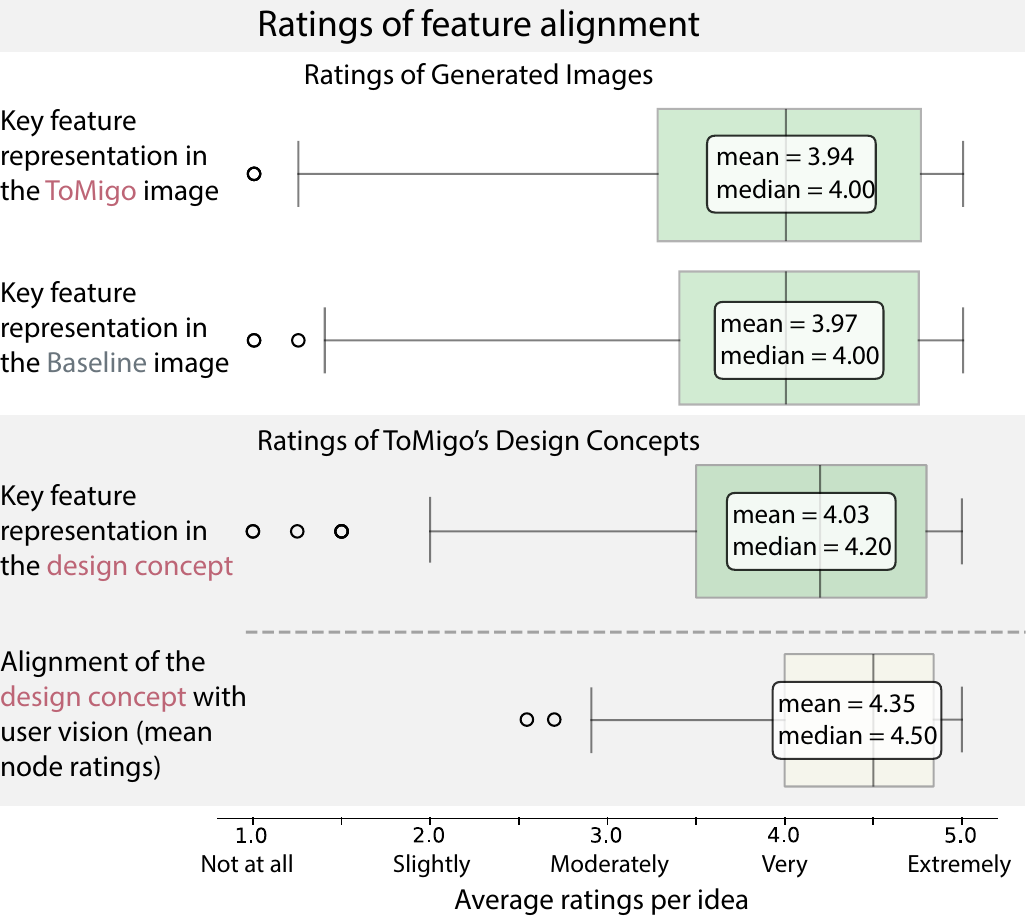}
    \caption{The box plots visualize ratings averaged per design idea. Users' evaluations of the two generated images based on their key features are similar, with medians of 4 (very well represented). The representation of these features in the concept graph was also high, with a median of 4.2. Average ratings for individual nodes were the highest among all categories, indicating that most nodes in ToMigo's graph align closely with user design ideas.}
    \label{fig:feature-ratings}
    \Description{Boxplots of user ratings of designs and graph concepts for key-features. ToMigo and Baseline designs show similar key-feature ratings, both with medians of 4.00 and means near 3.95. Graph concept ratings are higher, with key-feature ratings at median 4.20 (mean 4.03) and node ratings at median 4.50 (mean 4.35). Outliers appear at the lower end in all conditions.}
\end{figure}

\emph{RQ1:} When presented with the information in the graph nodes, participants rated its overall alignment with their original vision 4.22 on average (std: 0.77, median: 4) on the Likert scale, i.e. slightly more than \emph{very} well aligned. Figure \ref{fig:overall-ratings} shows the distribution of ratings in the lower subplot. Participants' ratings of how well the nodes represented their key features was 4.03 on average (std: 0.89, median: 4.2) and the ratings of alignment of individual nodes with the design vision were 4.35 on average (std: 0.57, median: 4.5). See Figure \ref{fig:feature-ratings} for a summary of the distributions. We can conclude that for most design ideas, the concept graph aligned very well with the design vision of the users.  

\emph{RQ2:} Regarding the impact of ToMigo on the generated designs, the mean overall alignment rating was 3.98 for ToMigo (std: 0.96, median: 4) and for the baseline 4.02 (std: 0.97, median: 4), indicating that both systems generated well-aligned designs. A mixed-effects TOST indicated equivalence within a 0.15 margin (p-values: lower=0.001277, upper=0.04596). Similarly, the average per-key-feature ratings were equivalent (0.15 margin, p-values: lower=0.003212, upper= $3.025 \times 10^{-5}$) with medians and means around 4 (ToMigo mean=3.94, std=0.93, median=4 and baseline mean=3.97, std=0.90, median=4). See Figures \ref{fig:overall-ratings} and \ref{fig:feature-ratings} for visualizations of overall and key-feature ratings respectively. We can conclude that ToMigo does not negatively impact the alignment of generated designs compared to the baseline.

We further analyzed the distribution of key-feature ratings and found that they exhibited greater variations than overall ratings, with standard deviations higher by 33\% for the baseline (1.25 vs. 0.94), 32\% for ToMigo (1.29 vs. 0.98), and 68\% for the graph (1.19 vs. 0.71). To better understand this discrepancy, we compared features against participants’ initial briefs and categorized them into three levels: explicitly mentioned, implied, and unrelated or contradictory. The majority of features were explicitly stated (49\%) or implied (35\%), indicating that most requests could be directly traced to participants’ descriptions. However, 16\% of features were contradictory, unrelated, or only weakly implied. These features often reflected considerations that participants had not initially articulated but became salient during the design process. 

This highlights how user intentions evolve dynamically: some requirements are fixed from the outset, while others emerge or shift as participants iteratively engage with design generation. These findings suggest that the theory-of-mind graph should not treat user intent as static but instead update its representation adaptively in interactions with users to capture their evolving design goals.  
\section{Applications}

We implemented multiple interactive features based on design concept graphs to demonstrate and test the benefit during a creative process. The features build on the intent prediction to improve alignment, making it transparent to support a grounding process between user and AI. The internal alignment and structure is used to afford reflection within the model and to encourage the user to structure their thinking about the design as well. The modularity of the graph is used to support iteration and refinement of designs.

\subsection{Theory of Mind Widgets}\label{tom-widgets}

\begin{figure*}
    \centering
    \includegraphics[width=\textwidth]{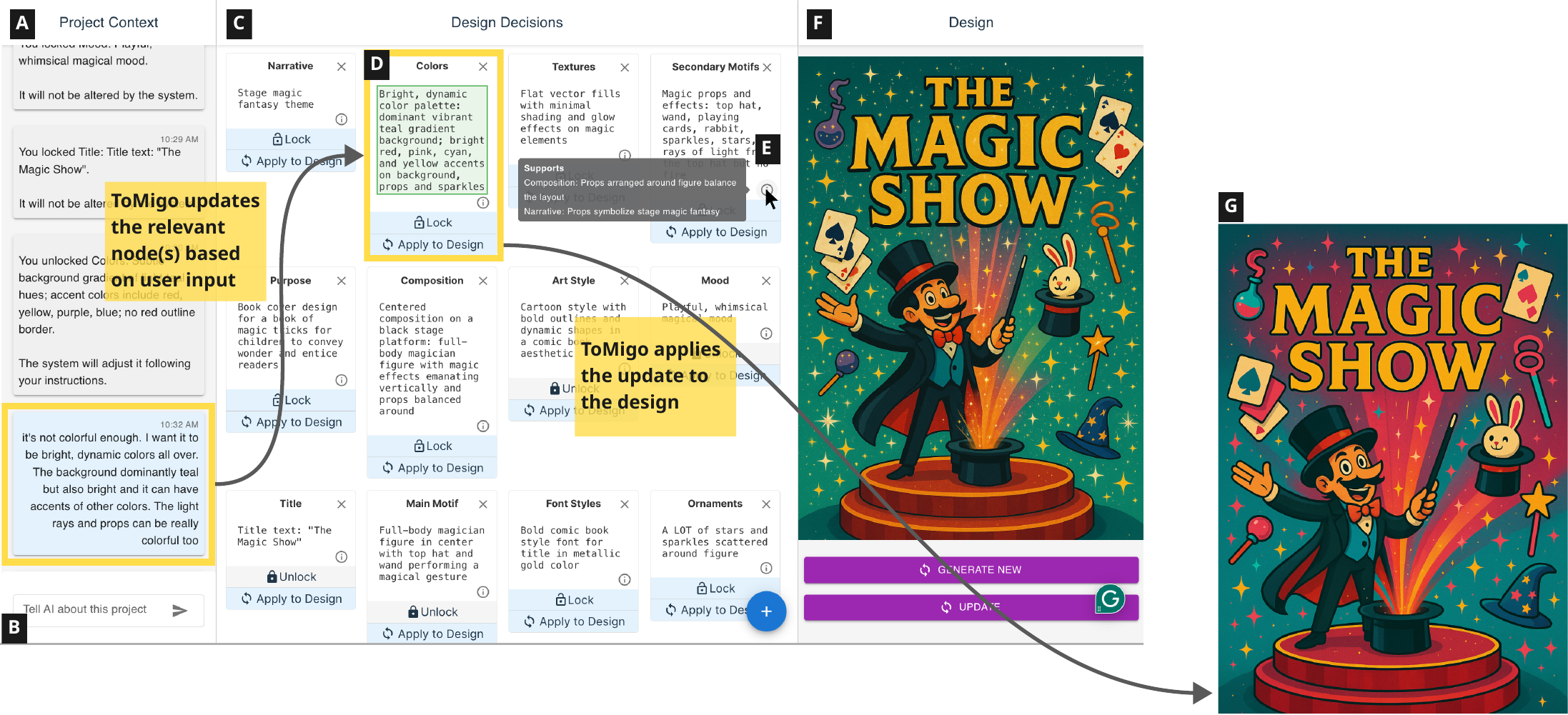}
    \caption{ToMigo application examples integrated into a UI. The project context panel (A) on the left contains the textual and visual user input and clarifying questions from ToMigo. Users can write messages to ToMigo (B), which is uses to update the nodes, shown as widgets on the design decisions panel (C). These provide a direct interface to design concept graph nodes allowing to inspect nodes (D) and edges (E) edit the node text, delete, add, lock to save and apply to design. The design panel (F) on the right shows the generated design and two options for generation: generate new uses the design concept graph to generate a new design, update compares the design to the current nodes and edits it to close the gap. Nodes with new content are highlighted in green until they are used for design generation. In this example, clicking ``Apply to Design'' causes ToMigo to emphasize the description of this node (D) in the existing design resulting in a realigned design (G).}
    \label{fig:placeholder}
    \Description{The ToMigo interface facilitates iterative design by connecting conversational feedback to a structured concept graph. The system displays the project history in the Project Context panel where the user specifies visual goals like a bright, dynamic color palette for "The Magic Show". ToMigo interprets these qualitative requests and automatically updates the relevant nodes in the Design Decisions panel. For example, the system modifies the Colors node to define a vibrant teal gradient background with red and yellow accents to match the user's intent. The interface also reveals the underlying logic of the design through visible edges that explain how Secondary Motifs like playing cards and sparkles support the overall composition and narrative. Once the user reviews these changes, they trigger the generation process to produce a final design that features a magician and a rabbit on a colorful stage. The resulting book cover directly reflects the updated node properties, such as the bold comic book font for the title and the dense arrangement of star ornaments.
    }
\end{figure*}

Theory of mind widgets directly build on the graph representation by making each node visible and editable in a widget. They provide a direct interface to view and manipulate the graph. Organized on a panel, they display the design decisions that combine into the full design concept. By showing the nodes with their types, they provide an overview to knowledgeable users and a framework for graphic design for novices. The node description is shown below and can be directly edited by the user. An information icon allows users to access the descriptions of the edges linked with that node, which provides insight into the internal alignment and reasoning. Users can also remove nodes, which will also remove their edges. They can also add nodes which then will be integrated into the graph with edges during the next global update. The widgets can be locked if users are fully satisfied with them, indicating to ToMigo to treat this node as confirmed by the user and not to alter it. 

To allow fine granular editing of the design based on the node interface, each widget can be applied to the design individually. ToMigo will then analyze how the feature can be realized or emphasized in the design and instructs an image-to-image model (GPT image) to apply the change.

\subsection{Clarifying Questions}
ToMigo uses the graph structure to generate questions that aim to clarify the design intent of the user to both the user and the system. Using the input history, graph, and schema it targets topics from the schema that have not been addressed yet in the input history. We developed question guidelines that build on the graph and schema with the following goals: (1) Bridging between the roles of design decisions (purpose, concepts, content, style), simultaneously gathering information about the user's view on nodes and edges to align to. (2) Fostering reflection on how to express turning communicative goals and high level concept decisions into visuals. (3) Using topics that were not yet discussed recently to encourage the user to think about different kinds of design decisions and make questions more interesting for the user. (4) Providing general guidance such as conciseness, improving accessibility for novices while being open to encourage users to reflect on their intention rather than leading them.

\subsection{Design Generation and Realignment}
We implemented three options for generating designs from concept graphs. The first, \emph{generate new} selects those nodes from the graph that are visual, converts them to a textual textual prompt and passes them to an image generator (GPT image) alongside the inspirational images. 

If the the design concept graph was updated, two options exist to \emph{realign} the design with the graph. These functions utilize the modular structure of the graph to prompt targeted changes using image-to-image generation (GPT image). The \emph{update} option compares analyses the gap between each node and the design and creates a shortlist of required changes to align the design more closely with the graph. If the user identifies a specific node that they want to introduce or emphasize more in the design, can specifically request to apply it (see \ref{tom-widgets}).

\subsection{System Implementation}
The system was implemented as a web application with a Python Flask backend, a PostgreSQL database, and a JavaScript React frontend. All API calls were handled through the OpenAI Responses API, with GPT-o4-mini (o4-mini-2025-04-16) serving as the basis for the ToMigo features described above. Image generation relied on GPT-4.1 (gpt-4.1-2025-04-14), which was the most recent version available at the time of completing the implementation. The full set of prompts used in the development process is provided in the supplementary materials.

\section{User Study 2}
We conducted a second user study to evaluate the effect of ToMigo enabled interactions on the process of generating designs and alignment with user intentions over time. The process of creation itself impacts the ability to express, and develop design intentions. Thus, we want to investigate not only the effect on the process itself but in extension on the development and communication of users' design intent. This study was guided by three research questions:

\begin{itemize}
    \item [RQ3:] Does the interaction with ToMigo improve the alignment with users' design intent? 
    \item [RQ4:] Does ToMigo support users in developing their design intent? 
    \item [RQ5:] Does ToMigo support users in communicating their design intent more effectively?
\end{itemize}


\subsection{Participants}

\begin{table}
\begin{tabular}{|p{5cm}|l|}
\hline
Experience                                                                                            & Count of responses                                                                                            \\ \hline
How often do you use generative AI to create or edit images (e.g., GPT, DALL·E, Firefly, MidJourney)? & \begin{tabular}[c]{@{}l@{}}Always: 1\\ Often: 4\\ Sometimes: 9\\ Rarely: 14\\ Never: 3\end{tabular}           \\ \hline
How interested are you in design?                                                                     & \begin{tabular}[c]{@{}l@{}}Extremely: 4\\ Very: 13\\ Moderately: 8\\ Slightly: 6\\ Not at all: 0\end{tabular} \\ \hline
How often do you create visual designs?                                                               & \begin{tabular}[c]{@{}l@{}}Always: 1\\ Often: 6\\ Sometimes: 13\\ Rarely: 8\\ Never: 3\end{tabular}           \\ \hline
\end{tabular}
\caption{Participants' experience with image generation and visual design, as well as interest in design.}\label{tab:participantinfo}
\label{tab:experience}
\end{table}

We recruited 32 participants (17 female, 15 male) excluding one from analysis who contradicted themselves in the questionnaire and interview. Their age range was 19 to 48 years, with a mean of 29.3. More information about the frequency of AI use for image generation, design interest and frequency of design creation can be seen in Table \ref{tab:experience}.

We reached participants via flyers and messages that we posted to local social media groups. The recruitment text mentioned that participants can pick a design task to generate with AI, aiming at primarily recruiting participants with a natural intention to generate a design. Participants were compensated for their time with a 12 Euro voucher to a local restaurant.

\subsection{Baseline}
As a baseline we implemented a system with a comparable interface to ToMigo but without the graph-based functionality. This is comparable with ChatGPT -- providing a chat functionality in the Context panel of the ToMigo interface. All user input in this chat was directly sent to GPT 4.1 (verison gpt-4.1-2025-04-14) with the image-tool option enabled, so that the generated response could contain text and/or images depending on the input. Text responses were displayed in the chat panel, while image responses were displayed in the design panel of the ToMigo interface.

\subsection{Materials}
Participants received an information sheet with details about the study condition, along with a consent form. A slide deck was used to explain the purpose, setup, and procedure of the study and to introduce the system’s features. To allow participants to test the system’s functionality, a practice design task was prepared with images and a brief. Both the ToMigo and baseline systems were made available as web applications on a laptop.

Participants also completed a post-task questionnaire, which included items on prior experience and interest in AI and design, as well as demographic data. As no standard measure exists for assessing AI alignment with user intentions, we developed a Gulf of Envisioning Questionnaire to assess the three gaps that can occur when generative AI is used in place of a design process \cite{subramonyam2024bridging}. The questionnaire included one pair of statements for each gap—the capability gap, the instruction gap, and the intentionality gap—as well as for envisioning overall. Each pair consisted of a positive valence statement about the system’s support in bridging the gap and a negative valence statement about the potential negative effects of the gap.

Finally, an interview script was used to gather qualitative insights into participants’ experiences with the systems. The interview began with an open-ended question inviting participants to describe their design process. This was followed by more specific questions about the role of the system in helping them pursue their design goals, its impact on their design intentions and communication, and their use of the different panels in the interface.

\subsection{Design}
Due to expected asymmetric learning effects we chose a between-subject-design. The ToMigo system would expose participants to the graphic design schema which they could carry over to the baseline condition to guide their design generation. 

\subsection{Procedure}
To encourage more natural user intentions and goal oriented design process, we asked participants to imagine a graphic design that they would like to generate and gather four or more inspirational images before arriving and submit them using a form so we could download them to the lab computer before the beginning of the study. To allows for a wide range of realistic design ideas, the instructions left open which kind of graphic design but to fit the scope of image generation, we specified that it would need to fit on one page. We encouraged participants to contact us if they were unsure whether their choice was a graphic design. 

Once the participant arrived, they were welcomed and introduced to the study purpose and procedure by the experimenter. In addition, they were given time to read a detailed information sheet before providing consent to the study conditions. The experimenter, then introduced the system the participant was going to use, before opening the system with a practice task. Participants could explore the functionality of the system until they indicated that they were ready to proceed to the main task. For the main task, participants used the system for up to 15 minutes without interruption or until they indicated that they were satisfied with the design. After completing the main task, they filled the questionnaire and were interviewed by the experimenter.

\subsection{Data Collection}
We recorded the following data: Text and image input to the system, generated designs, logs of all interaction with the system with time stamps, screen recording during the the main task, questionnaire answers and audio of the interview.

\subsection{Analysis}
The interview data was transcribed and thematic analysis conducted for themes related to the research questions. Mann-Whitney U tests were run to compare each measure in the Gulf of Envisioning Questionnaire comparing baseline and ToMigo. 

\subsection{Quantitative Results}
\begin{figure*}
    \centering
    \includegraphics[width=\textwidth]{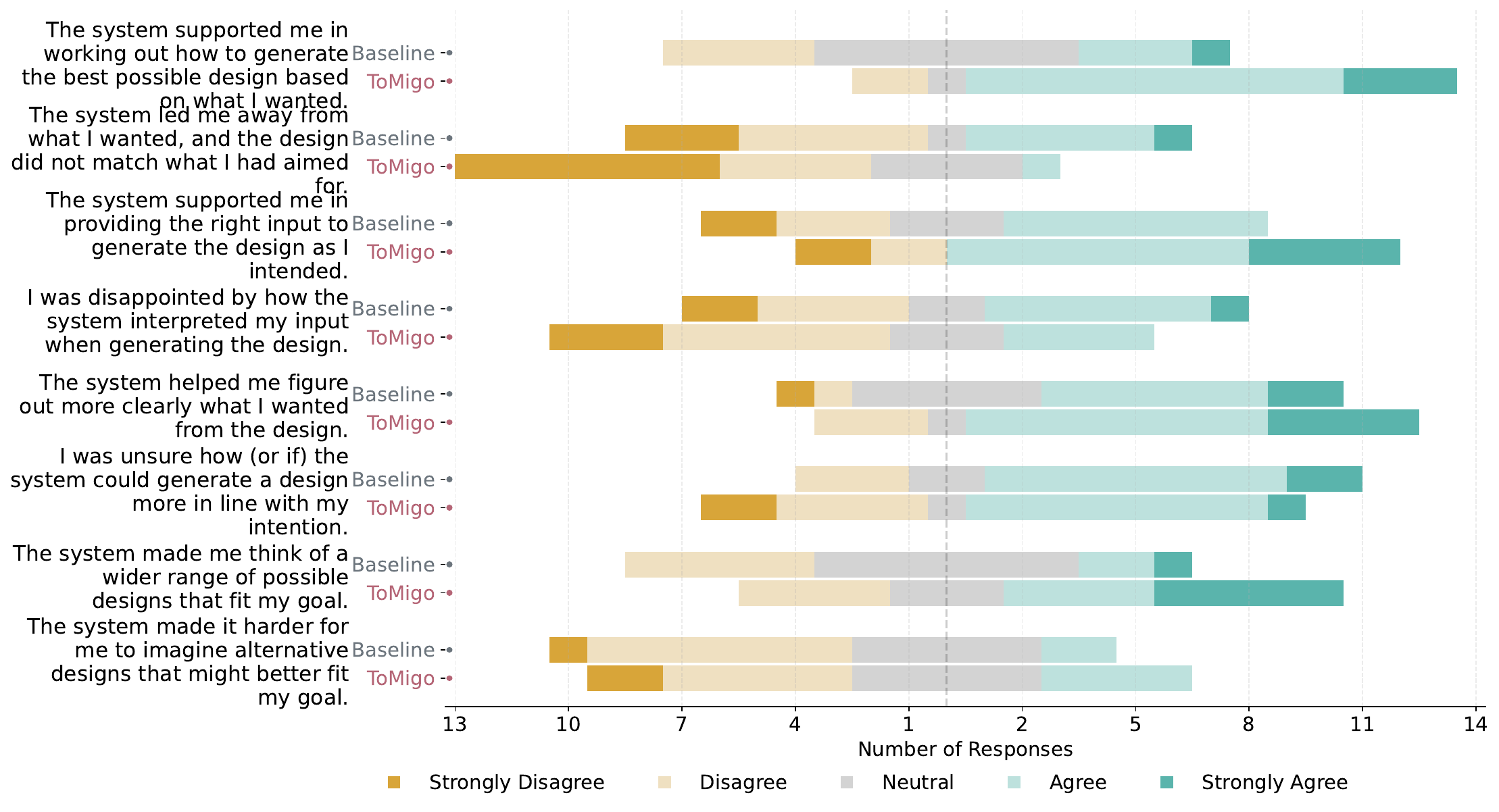}
    \caption{Participants' agreement ratings with the statements in the questionnaire. In particular the first two questions targeting the gap of capability indicated improvements through ToMigo, which were significant. }
    \label{fig:likert}
    \Description{Stacked bar charts display participants’ agreement ratings for eight questionnaire statements, comparing a Baseline system with ToMigo. Each statement features two horizontal bars where responses range from "Strongly Disagree" (left/orange) to "Strongly Agree" (right/teal).

Generating the best design: ToMigo shows a significant shift toward "Strongly Agree" compared to Baseline, which has a larger "Disagree" segment.

Leading away from goals: ToMigo received more "Strongly Disagree" responses than Baseline, indicating users felt it stayed more on track.

Providing the right input: ToMigo shows higher "Strongly Agree" counts, while the Baseline ratings cluster more around the "Neutral" and "Agree" segments.

Disappointment with interpretation: ToMigo has a notably larger "Strongly Disagree" section than Baseline, suggesting higher satisfaction with how the system processed inputs.

Clarifying design goals: ToMigo responses show a heavy concentration in the "Agree" and "Strongly Agree" categories, outperforming the more neutral Baseline results.

Confidence in alignment: ToMigo shows stronger agreement that the system could generate designs matching user intention, whereas Baseline shows more "Disagree" responses.

Thinking of wider designs: ToMigo prompted more "Strongly Agree" ratings for exploration, while Baseline leaned more toward "Disagree" and "Neutral."

Difficulty imagining alternatives: Both systems show a spread of disagreement, but ToMigo has fewer "Agree" responses than Baseline, indicating it hindered users less.

Overall, the ToMigo bars consistently extend further into the teal "Agree" and "Strongly Agree" zones for positive statements and further into the orange "Disagree" zones for negative statements.
    }
\end{figure*}

For the questions targeting the gap of capability, which is about understanding the necessary steps to reach the design goal, ToMigo received significantly better ratings. Participants agreed significantly more that ToMigo supported them in working out how to generate the best possible design than the baseline (ToMigo mean=3.88, baseline mean=3.07, U=60, p=0.01). The differences in the other scales were not statistically significant. Figure \ref{fig:likert} visualizes the results of the questionnaire.

\subsection{Qualitative Results}

\subsubsection{ToMigo provided grounding opportunities.}
Participants in the ToMigo condition frequently mentioned themes around checking the AI’s understanding through the widgets. P2 emphasized how this process clarified the system’s reasoning: ``The widgets have a lot of categories, and I can see which ones are changed [..] I think it helped me understand how the model works.'' This grounding gave them confidence to proceed — ``I saw the AI change some widgets, and I knew exactly what it changed. I read the changes and thought, well, it’s almost perfect, it expressed my idea.'' P15 also valued this interplay, describing ToMigo as ``letting me preview rough concepts and ideas, and then edit them,'' while writing to the chat interface when uncertain: ``I said, ‘the blue doesn’t fit the vibe,’ without specifying other colors. I let the system figure it out.'' For P5, the design decisions panel worked as a validation tool: ``I used it like a sticky note. I confirmed from the context panel that the AI actually got what I meant — okay, it got this part right, it got that part right.'' By contrast, participants who only had the chat interface grounded understanding more loosely. P9 noted, ``I relied more on the chat panel'' in contrast to generating via button press, ``because I would just give feedback,'' and P20 described reacting directly to outputs: ``With different pictures I had different ideas in mind.'' However, other participants noted the messages from the baseline system were not useful for grounding as they were misaligned: ``I received two kinds of feedback in the form of the generated image, and also a comment in the chat, what the system had changed. And they were often not aligned.'' (P18) Whereas baseline participants grounded through conversational adjustments, ToMigo participants grounded by inspecting and confirming how the AI interpreted their intentions before moving forward. 

\subsubsection{ToMigo helped discover design requirements.}
Both widgets and clarifying questions helped participants sharpen their design intentions without leading them away from their goal. Several participants described how ToMigo supported them in discovering design decisions more clearly. P2 explained that the system made their idea more obvious and allowed them to reach a good outcome. For P4, ToMigo helped to develop their descriptions into richer design directions: ``It makes my intention or my problem, my description more detailed in the motif or the general idea or the general feeling of the poster, of the image.'' Similarly, P11 emphasized that their overall design intention did not change but became ``more concrete and more new.'' P15 illustrated this balance, noting that ToMigo sharpened details about the composition of their poster without distracting from the broader vision: ``I think I had a clear idea of how I wanted the poster to look beforehand. So it was more the details that I figured out with this, not the overall concept.''  

Beyond clarifying intentions, widgets and questions also guided participants toward aspects of design they might not otherwise have considered. P4 described how the questions ``helped me to notice features I hadn’t thought about before.'' P15 similarly reflected, ``I thought the ideas that came up were very good, especially things like colors and textures. These are things I wouldn’t have thought about or prompted necessarily,'' and continued, ``It helped me figure out what exactly I wanted by suggesting certain design decisions I was not aware of beforehand.'' P6, who had limited design knowledge, highlighted how the design decisions panel expanded their awareness: ``It let me know the different aspects of the design — mood, colors, the key elements, and ornaments — which help the design enhance or convey the message.'' Reflecting on the same panel, they added: ``I specifically went there and kept my intent there. If there had been no separate design decision called ,,'art style,'' I might not have thought about that aspect. It clearly improved my intention for the design and helped me in a positive way for creating it.'' In this way, ToMigo actively prompted attention to dimensions of design that users might not otherwise have noticed.  

By contrast, participants in the baseline condition reported that feedback remained repetitive and focused only on broad aspects. As P8 put it, ``The messages were more like the same thing. If I wanted to change some icons or some colors, I could, but it didn’t go beyond icons and colors in layouts in general.'' This contrast illustrates how ToMigo’s widgets and clarifying questions expanded participants’ design space, while baseline feedback reinforced existing ideas without opening new directions.  

\subsubsection{ToMigo supported reflecting on the design vision}
Participants in the ToMigo condition often described how the system supported reflection on their overall design vision, both by helping them connect individual decisions into a coherent whole. P4 highlighted the edge inspection tooltips as particularly valuable for seeing how decisions were interrelated: ``it was very clear that, okay, this supports the color or this supports the motif, so that helps with the general vision of the image.'' Reflection on the goals of the design was also encouraged by clarifying questions. For P4, being asked questions directly in the system ``made me think about what I want in the photo [part of their design], which I didn’t think of before.'' P11 similarly emphasized that ``the most impressive thing for me is it gave me the hint or the motivation about what I want.'' P6 explained how receiving these questions sharpened their intent: ``When I got the questions back from the AI, then my intentions got crisper and clearer and narrower. I think that is one of the aspects where my intent got clearer and better.'' P5 described this process as a way of clarifying what to include and exclude: ``I would say it just clears me out of what I don’t want and what I do want more.'' P23 also noted that the questions helped her develop her design further: ``They do influence my ability to design, maybe making me pay more attention to the details.''  

By contrast, participants in the baseline condition primarily reflected on the generated visuals themselves and adjusted details from there. P8 noted, ``I think it gave me a good starting point. If I was going to do this by myself, I think it would take me a lot longer to get to this.'' P9 described the system as more of a production assistant than a co-designer: ``It’s painter. I think I’m the designer, the system is just helping me draw the poster. It’s my assistant.'' At the same time, P9 reported significantly altering their intention for the better after encountering the first generated design: ``I realized, so on the beach, a bar is not very beautiful, not realistic [..] So I accepted the system’s design.'' They later reflected that ``finally my intention became expressing a relaxing atmosphere.'' P13 similarly explained that the generated images ``helped me visualize how it could be put in other ways, and I also realized that maybe I could go a bit more in depth in describing how to prompt it.''  

Taken together, these accounts suggest that ToMigo encouraged reflection by making design goals explicit, linking them to supporting elements, and prompting sharper articulation through questions, whereas the baseline supported reflection more indirectly through reactions to the outputs.  

\emph{ToMigo as a domain fluency boost.}  
Participants frequently described ToMigo as translating vague or inarticulate ideas into concrete design cues. P4 emphasized that ``the best feature of the system is that it can translate my vague intentions into cues [widgets],'' noting that it helped carry intentions all the way ``to the actual result of the poster.'' Alongside translation, participants highlighted how ToMigo supported their fluency in expressing design ideas. P1 reflected that the system pushed them to be more articulate about the ideas in their head. For P11, the design decision panel felt ``more accessible'' than general tools like ChatGPT, while P17 appreciated that even without technical terms, the system could recommend and explain them in a way that matched their intent. P15 illustrated this dynamic, noting that by editing the design decisions already formulated by the system they were ``more fluent in expressing my ideas'' than if they had been given ``empty boxes'' to fill. They explained that prompts like ``composition'' would not have led them to detailed concepts such as ``layered composition'' or ``foreground motif'' without this guidance. Other participants echoed this sense of support: P6 described it as giving a ``normal user…the ability to tweak key design aspects,'' and P5 emphasized how the wording helped them compensate for their lack of background in color theory, ``picking out the better way to describe what I want.'' In these ways, ToMigo acted both as a translator of intention and as a scaffold for developing fluency in design vocabulary and expression.

\emph{Direct manipulation and modularity}  
Participants valued how ToMigo allowed them to directly manipulate design intentions through modular widgets. P6 described modifying design intention widgets as ``one of the best things'' they had experienced compared to other design aids, emphasizing that ``the design decisions felt like I had control over the design, what I am designing.'' This sense of control also came from being able to focus selectively on different aspects. P21, for example, explained that they locked the colors once satisfied and then ``mostly changed or tried to change the details and content of the poster.'' P5 illustrated this fine-grained control in practice: they could ``delete or change the wording from vivid to desaturated or small subtle'' when the output felt too bright, and adjust phrasing to achieve ``blurred outlines'' or other specific effects.  

In contrast, participants in the baseline condition reported less modular control. P10 admitted they avoided clicking the ``generate new'' button out of fear it would change everything, relying instead on chat instructions to make edits. They noted they would have welcomed the possibility of editing only one design aspect at a time. Similarly, P13 reflected that the baseline system ``made me realize that maybe it is not able to make too many fine details such as a blade,'' pointing to the limitations of working without explicit modular handles. Together, these accounts highlight how modularity in ToMigo supported both confidence in directing the AI’s output.  

\subsubsection{Overwhelming amount of widgets}
A few participants (P1 and P21) complained that the amount of widget displayed overwhelmed them. P1 decided to turn to the chat interface for this reason later contemplating that reading and editing the widget could have been easier. Referring to the time limit of the task, P12 thought that ``with more time, I think I could have read them all and then changed the details from the boxes [widgets], but now it felt easier just to tell the chat what I want.''

\section{Discussion}
Our user studies validate that ToMigo fulfills its design considerations (Section \ref{design-considerations}) by supporting users through the co-evolution of design problems and solutions.

Study 1 demonstrates that our theory of mind model effectively aligns with user intentions, satisfying DC1. This study quantified alignment with initial user goals as a foundation for interactive refinement. Participants used Likert scales to rate how well the predicted intentions and generated designs matched their vision. The high alignment ratings for both user specified requirements and system inferred ones confirm that the system captures both explicit and implied features from sparse inputs.

By filling in necessary but unspecified design attributes, the system maintains internal coherence as required by DC2. This automatic gap-filling provides essential support for novices who were reported discovering design requirements with ToMigo. Study 2 highlights how the interpretable and modular nature of the graph facilitates creative growth. The structured representation provides an inspectable "boundary object" that fulfills DC3 by making the AI's reasoning transparent. Participants used the edge tooltips and widgets to understand how individual decisions supported their broader goals. This transparency encouraged users to reflect on and sharpen their intentions.

The modularity of the graph further enables the iterative refinement described in DC4. Users successfully evolved their ideas by making fine-grained adjustments to individual nodes without losing control of the overall concept. Finally, the system achieved grounded generation as specified in DC5 by realigning the final designs with these updated graphs. Participants reported that this direct connection between the conceptual graph and the visual output provided a sense of control that was absent in baseline chat interactions.

\section{Limitations and Future Work}
We explored different visual design tasks in our experiments, these were, however, limited to one page static graphic design types. They, therefore, did not include more complex design types like multiple page set-ups, interactive user interfaces, motion graphics, or multiple sides of three dimensional packages (where we had studied only the rendering of one side). While many of these aspects could be modeled through our existing graph schema, an extension of the schema might be necessary, e.g., to cover interactive elements. 

The information graph in this work is based on inspirational aspects and their relations expressed in images selected by non-designers since our goal was to model what internet users consider inspirational. This might be different from professional designers' use of reference images. The relations might also not be complete in terms of established design knowledge. The schema should be considered a starting point for reference based design similar to a prior for user intentions that avoids a cold start. It should be handled flexibly not as a complete schema or guideline of a professional design. 

\section{Conclusion}
This work introduced ToMigo, a system that represents user intentions as design concept graphs. It leverages these graphs to align AI understanding with creative goals. Our contributions include a graph schema for design concepts and computational methods to construct and refine aligned design concept graphs from simple verbal and visual inputs. We also contribute three interaction techniques: theory of mind widgets, clarifying questions, and graph-based design generation. These techniques demonstrate how graphs make AI reasoning interpretable and controllable.

The studies showed that this approach helps users clarify and evolve their intentions while preserving control. Theory of mind widgets supported inspection and editing of the AI’s understanding, clarifying questions prompted deeper reflection on goals, and design generation aligned with the graph allowed outputs to remain coherent with adapted design concepts. Compared to the baseline, participants using ToMigo reported greater ability to confirm, adjust, and expand their ideas without losing sight of the larger design vision. ToMigo participants reported that the system helped them translate vague ideas into concrete cues and discover new aspects like textures or colors. In contrast, baseline feedback was described as repetitive and limited to broad aspects.

In answering our research questions, we conclude that design concept graphs bridge the gap between simple prompting and complex creative practice. They achieve this by connecting high-level purpose to low-level features in a structure that remains modular and editable, enabling AI to stay aligned with evolving user intent.

\begin{acks}
This work was supported by the Finnish Foundation for Technology Promotion and the Finnish Center for Artificial Intelligence. Experiments involving API calls to GPT 4.1 and GPT o4 mini were partially supported by a compute grant from OpenAI.
\end{acks}

\bibliographystyle{ACM-Reference-Format}
\bibliography{99_refs}

\end{document}